# Spatial Registration Evaluation of [$^{18}$F]-MK6240 PET

James Zou[1], Aubrey Johnson[1], Jeanelle France[1], Srinidhi Bharadwaj[1], Zeljko Tomljanovic[1] M.D., Yaakov Stern[1] Ph.D, Adam M. Brickman[1] Ph.D, Devangere P. Devanand[1,4] M.D., José A. Luchsinger[2,3] M.D., William C. Kreisl[1] M.D., and Frank A. Provenzano[1] Ph.D

1. Taub Institute, Columbia University Irving Medical Center, New York, NY, USA
2. Department of Medicine, Columbia University Irving Medical Center, New York, NY, USA
3. Department of Epidemiology, Columbia University Irving Medical Center, New York, NY, USA
4. New York State Psychiatric Institute and Department of Psychiatry, Columbia University Irving Medical Center.

## Abstract:

Image registration is an important preprocessing step in neuroimaging which allows for the matching of anatomical and functional information between modalities and subjects. The most commonly used image registration algorithms today take advantage of similarities between image signal intensities to find a corresponding match. This can be challenging if there are gross differences in image geometry or in signal intensity, such as in the case of some molecular PET radioligands, where control subjects display relative lack of signal relative to noise within intracranial regions, and may have off target binding that may be confused as other regions, and may vary depending on subject. This can cause inappropriate image warping or even obvious registration failure which then require manual adjustments, slowing analysis and introducing variance. The use of intermediary images or volumes have been shown to aide registration in such cases.

To account for this phenomena within our own longitudinal aging cohort, who are non-demented subjects with no obvious tau-PET pathology, we generated a population specific MRI and PET template from a broad distribution of 30 amyloid negative subjects. We then registered the PET image of each of these subjects, as well as a holdout set of thirty 'template-naive' subjects to their corresponding MRI images using the template image as an intermediate using three different sets of registration parameters and procedures. To evaluate the performance of both conventional registration and our method, we compared these to the registration of the attenuation CT (acquired at time of PET acquisition) to MRI as the reference. We then used our template to directly derive SUVR values without the use of MRI.

We found that conventional registration was comparable to an existing CT based standard, and there was no significant difference in errors collectively amongst all methods tested. In addition, there were no significant differences between existing and MR-less tau PET quantification methods. We conclude that a template-based method is a feasible alternative to, or salvage for, direct registration and MR-less quantification; and, may be preferred in cases where there is doubt about the similarity between two image modalities. Additionally, our method could be of use in a clinical setting when there is a need for broad analysis or when no secondary modality is available.

# Introduction:

Image registration, the spatial alignment of one image into another image's space, is an important preprocessing step for both group and single-subject neuroimaging analyses. It consists of estimating a transformation and aligning one image into another using an image's provided anatomy. Various algorithms and software suites are available today to help in the registration of images (*1*). While the registration of two images can be accomplished through various methods, the most common methods involve the matching images based on similarities in image intensity (*2*) at the voxel level, as they do not require prior segmentation of images and can be applied to an array of modalities. However, this assumes that there is adequate similarity in signal for a given structure present in both the reference and moving image, a precondition that is not always valid, especially when dealing across modalities where structural information (i.e. hydrogen density in MRI) may not necessarily correlate reliably with functional information (i.e. neurofibrillary tangle and off-target deposition in Tau PET) (*3*).

The recent development and use of positron emission tomography (PET) radioligands for the identification of tau tangles in vivo provides a new method to measure tau accumulation both topographically and quantitatively. The first studied ("first-generation") radioligands such as $^{18}$F-AV-1451 have been used to estimate regional increases in tau in longitudinal studies (*4,5*) and to define characteristic patterns of suspected tau burden of various tauopathic diseases (*6*). However, these radioligands are confounded by nonspecific binding to non-tau targets in as well as other non-brain (*7*) and extra-cranial areas. A "second-generation" radioligand, $^{18}$F-MK-6240, appears to have a reduced pattern off-target binding (*8*) while still maintaining regional binding patterns that agree with post-mortem tau topography in AD (*9*).

In the case of tau PET, for subjects with relatively low signal, the image co-registration is likely to be dominated by off-target or discordant anatomy (see Figure 1). This can be a potential issue because (a) the intensity and location of off target binding between subjects and even between chronological scans is not well characterized and (b) there may be intensity differences in corresponding regions in the same patients across modalities (e.g. the leptomeninges in $^{18}$F-MK6240 are relatively hyperintense while relatively hypointense on T1 MRI). This therefore may lead to errors in registration which may hamper the ability to accurately measure tracer uptake in small regions of the brain traditionally implicated in very early AD pathophysiology, such as the entorhinal cortex. Ensuring proper registration of PET tracer uptake onto anatomical data would aid the field in creating models for the early detection of pathology as well as monitoring the efficacy of early interventions.

The creation of 'template' images – averaged and registered images of an imaging modality created from a given study population – provide both a range of imaging characteristics across a given cohort and are useful tools for longitudinal analyses, especially when there is a need to account for disease characteristics (such as cortical atrophy) or for potentially poor quality intermodal (i.e. motion artifact in a patient's corresponding T1 MRI) reference images (*10*). Templates have been created for a wide variety of imaging modalities and tasks, including for the registration of images (*11–13*). Using an intermediary template has been shown to have utility in cases where there may be a geometric or large intensity difference between two images in the same subject, such as in the measurement of tumors for measuring progression or surgical/therapeutic planning; or in the case of degenerative diseases such as AD where there is potential for significant time during scans. These templates can then also be used to create atlases, which can be used for applications such as MR-less quantification of uptake (*14–16*), especially for subjects unable to obtain an MRI scan due to timing or other exclusionary criteria such as pacemakers or implants.

A template created for $^{18}$AV-1451 using older healthy controls (*17*) showed spatial tau deposition for subjects as a result of normal aging, as well as clarified off-target binding of the compound, and another method utilizing a template and machine learning was shown to have comparable performance in spatial normalization (*18*). While results with these templates generally agreed with other recent observations about tau pathology - it also reiterated the substantial choroid plexus binding, a particularly difficult region given its relative location to the MTL, which somewhat limits the usefulness of this template. While Betthauser et al. (*8*) have created an averaged template image for this compound for purposes of visualization, to our knowledge they have not used this template for registration. To our knowledge, neither conventional direct registration nor any template-based registration of $^{18}$F-MK6240 images have been evaluated extensively to determine whether off-target binding of this radioligand adversely affects alignment to MRI and definition of regions-of-interest

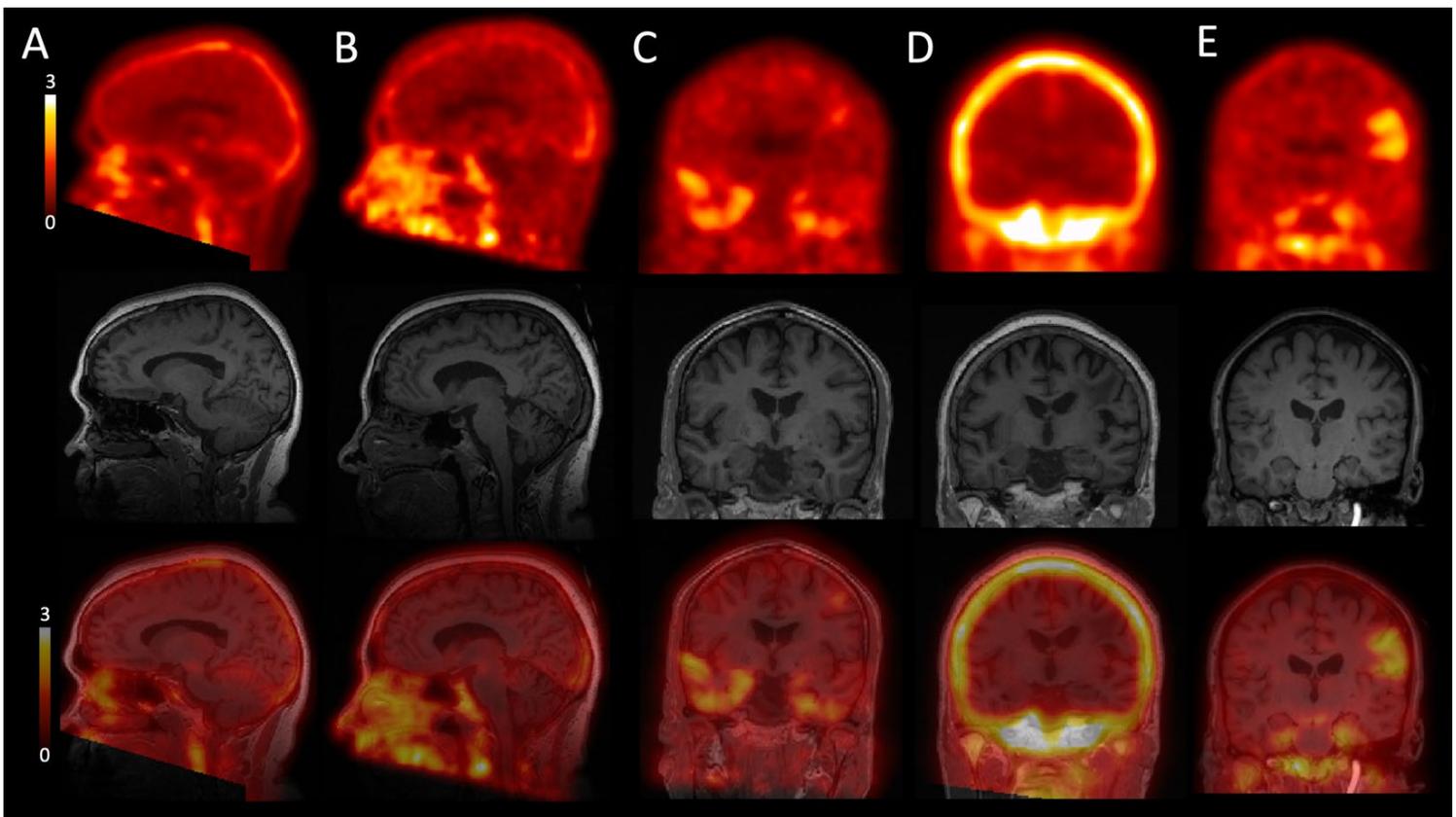

**Figure 1: Example subjects from the cohort partially illustrative of range of phenomena.** All subjects MK6240 images (top row) are in MRI space (middle row), and have been smoothed with a 3mm kernel for visualization purposes, and overlaid with 50% opacity (bottom row). The window for each PET image is consistent across subjects (0 - 3). A. 60F, Amyloid negative subject without significant tracer uptake discovered. B. 68F, Amyloid positive subject without significant tracer uptake discovered. C. 61M, Amyloid positive subject with significant tau discovered in the medial temporal lobes, and subsequently excluded from further analysis. D. 61F, Amyloid negative control with probable old infarction incidentally discovered on T1 MRI (image right), but with negative findings on MK6240. Note significant leptomeningeal binding in this subject. E. 65F, Amyloid negative control with incidental cortical tau discovered on imaging, without MRI correlates (T1 or FLAIR (not shown)) or clinical history/symptoms to suggest injury or underlying pathology.

## Methods

### Subject Recruitment

Participants were selected from the Northern Manhattan Study of Metabolism and Mind (NOMEM) research cohort at Columbia University Irving Medical Center (CUIMC). NOMEM is a longitudinal, community-based study of middle-aged, non-demented, Hispanic men and women. From this larger study, we selected 201 subjects who had $^{18}$F-MK6240 PET performed between 07/01/2018 and 07/01/2019. Inclusion criteria also included age 55 - 69 years, availability of clinical and neuropsychological assessments, 3T brain magnetic resonance imaging (MRI), and Positron Emission Tomography (PET) with injection of the amyloid radioligand $^{18}$F-Florbetaben. Exclusion criteria included a diagnosis of dementia and cancer other than non-melanoma skin cancer. The interval between amyloid PET and MRI was 4.17 ± 8.43 days. The interval between MRI and tau PET was 210 ± 86 days. This study was approved by the Institutional Review Board and the Joint Radiation Safety Commission at CUIMC. All study participants provided written informed consent. Funding sources had no role in study design, data collection, data analyses or interpretation

### Imaging

All subjects underwent a T1-weighted MRI sequence using the BRAVO protocol on a GE 3T MRI machine. An $^{18}$F-Florbetaben (injected activity = 8.1 mCi; images acquired 90-110 min post-injection) and $^{18}$F-MK6240 (mean injected activity = 4.48 mCi; images acquired 80-110 min post-injection) were each acquired on a Siemens Biograph mCT. For the purposes of this study we were also able to acquire 60 computed tomography (CT) scans acquired at the same time as the $^{18}$F-MK6240 emission scan.

### Image Processing

All MRIs were visually inspected (AJ) for notable artifact (e.g. motion, field of view cuts, etc.) as well as to confirm absence of chronic infarcts or other structural pathology. MRI scans were then automatically skull stripped using Freesurfer's watershed function (*19*) and then visually inspected for fidelity of extraction, correcting mislabeled voxels with connected cluster analysis using FSL's cluster function. Additionally, total intracranial volume (ICV) was calculated from this MRI using Freesurfer 6.0(*20*).

PET scans were visually inspected by an experienced reader (WCK) for determination of amyloid status, using previously validated criteria (*21,22*). In this sample of subjects without dementia, we excluded those with positive amyloid PET scan or visibly discernible tau signal. The included subjects therefore were expected to have minimal on-target tau signal with $^{18}$F-MK6240 PET.

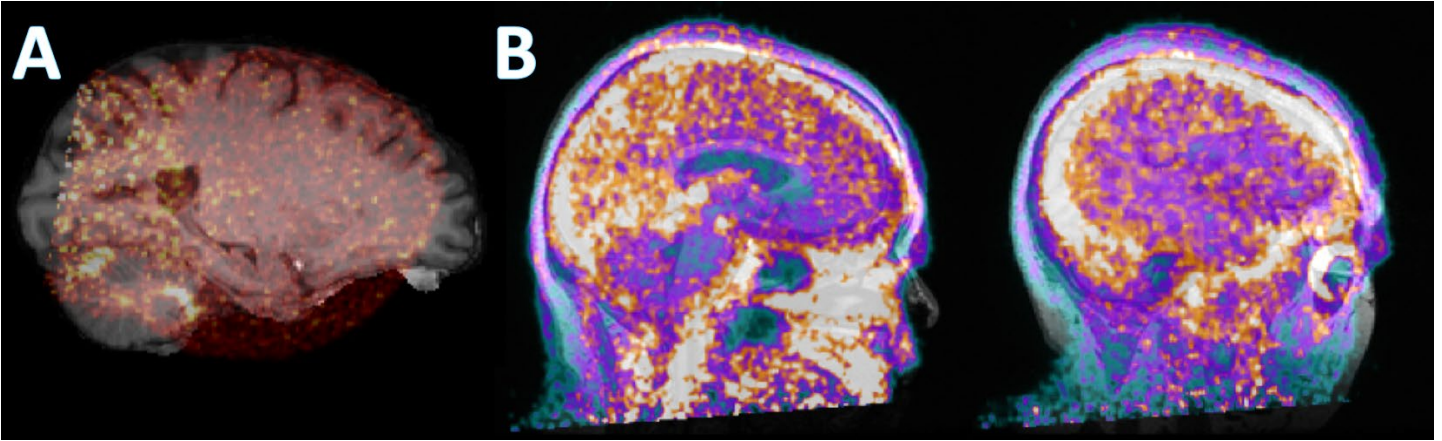

**Figure 2: Examples of attempted registration using (a) brain extracted MRI and (b) whole brain MRI as reference image in control subjects without significant on-target tau binding.** Note in (a) complete inability of standard registration algorithm to properly align input and reference images, and in (b) inappropriate matching of leptomeningeal signal to extracranial areas (including outside the subject).

**Template Creation**

We randomly chose 30 amyloid negative subjects among all controls whom collectively reflected the sample distribution in terms of age, gender, ethnicity and intracranial. We chose 30 subjects because we believed this to be an optimal number given considerations for accuracy/representativeness of template generation versus requirements for having holdout subjects for unbiased evaluation (*23*). Using FSL and the image registration suite ANTS (*24*,*25*), we created both an MRI and PET template from these 30 subjects scans, respectively (see Figure 3 for the finished template).

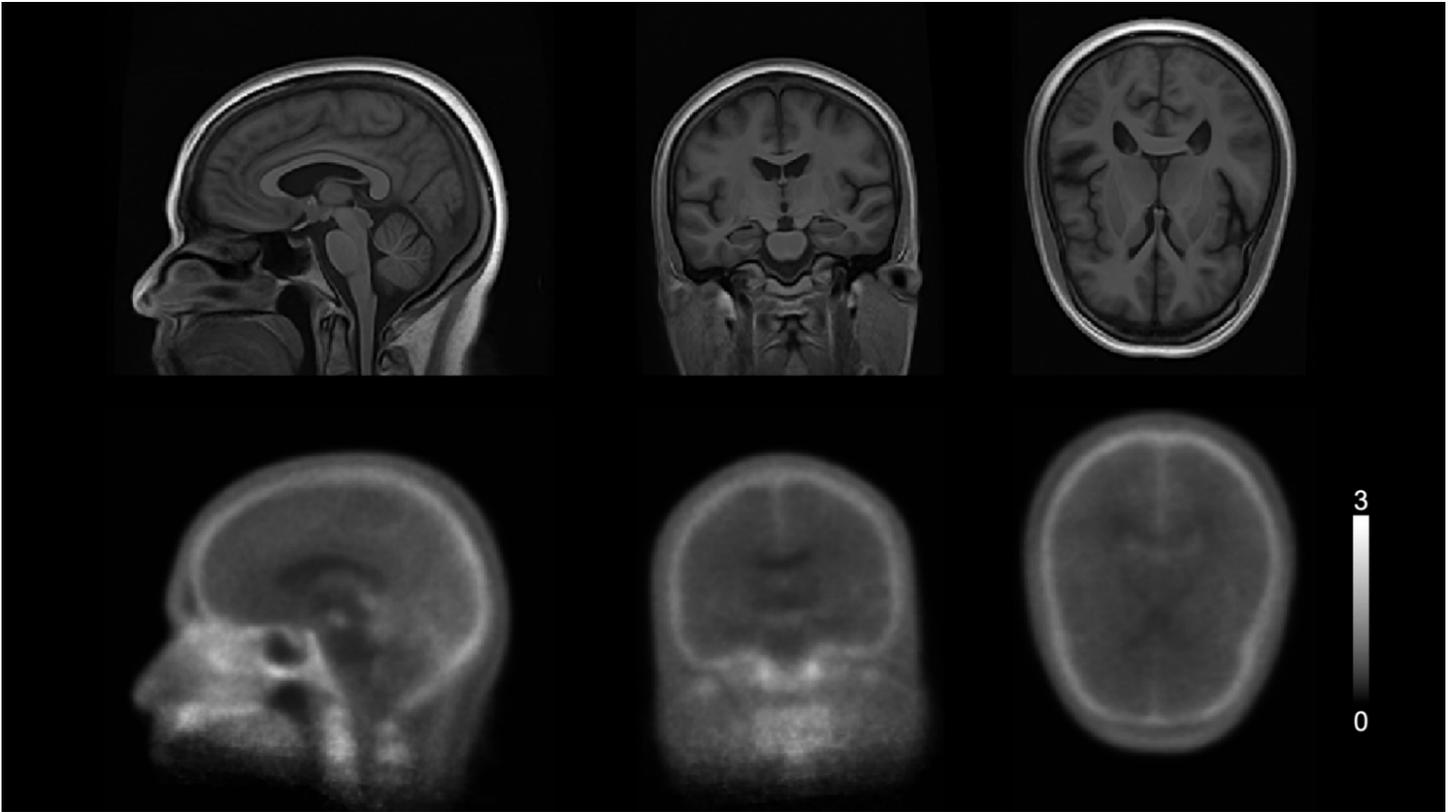

**Figure 3: Top: MRI template** created using ANTs from 30 selected subjects. **Bottom: PET template**, created using the PET registered to template MRI space using the attenuation CT as an aide.

**Registration Parameters**

For the registration procedures in this study, we utilized the open source software FLIRT (*26*,*27*). Parameters for each method of registration are fully listed below. For cost function we evaluated both mutual information (*3*) and correlation ratio (*28*).

| Method | Moving Image(s) | Reference Image(s) | Cost Function | Interpolation | Degrees of Freedom |
|---|---|---|---|---|---|
| CT aided | CT scan | Subject MRI | Mutual Information | Trilinear | 6 |
| Direct Registration [direct] | Subject PET* | Subject MRI | Mutual Information | Trilinear | 6 |
| Intermediate Template [Template_1] | Subject PET, Subject MRI | MRI template, PET template, Subject MRI | Mutual Information | Trilinear | 12 |
| Transformed Template (MI) [Tempalte_2] | Subject PET, PET template, MRI template | Subject MRI, transformed PET | Mutual Information | Trilinear | 6 |

| | | template, transformed MRI template | | | |
|---|---|---|---|---|---|
| Transformed Template (CR) [Template_3] | Subject PET, PET template, MRI template | Subject MRI, transformed PET template, transformed MRI template | Correlation Ratio | Trilinear | 6 |

**Table 1: Outline of registration parameters used.** Moving images refer to any image which had either registration calculated for or transformation applied to. The Subject PET used for each registration is the same PET image which has been first motion corrected then averaged from 90-110 minutes. The title in [brackets] is shorthand which we use in proceeding sections.

**Template based Registration Methods**

We tested the utility of using these intermediate templates for registration of subject PET into native MRI space (see Figure 4A). This method is referred to in our results as the 'intermediate' template (Template_1).

As an alternative, because we wished to avoid interpolation artifacts of transforming the PET image into template space (and avoid introducing any uncertainty due to excess degrees of freedom on the subject PET image), we also performed registered by transforming the templates instead (See Figure 4B). This method is referred to in our results as the 'transformed' template (Template_2 and Template_3).

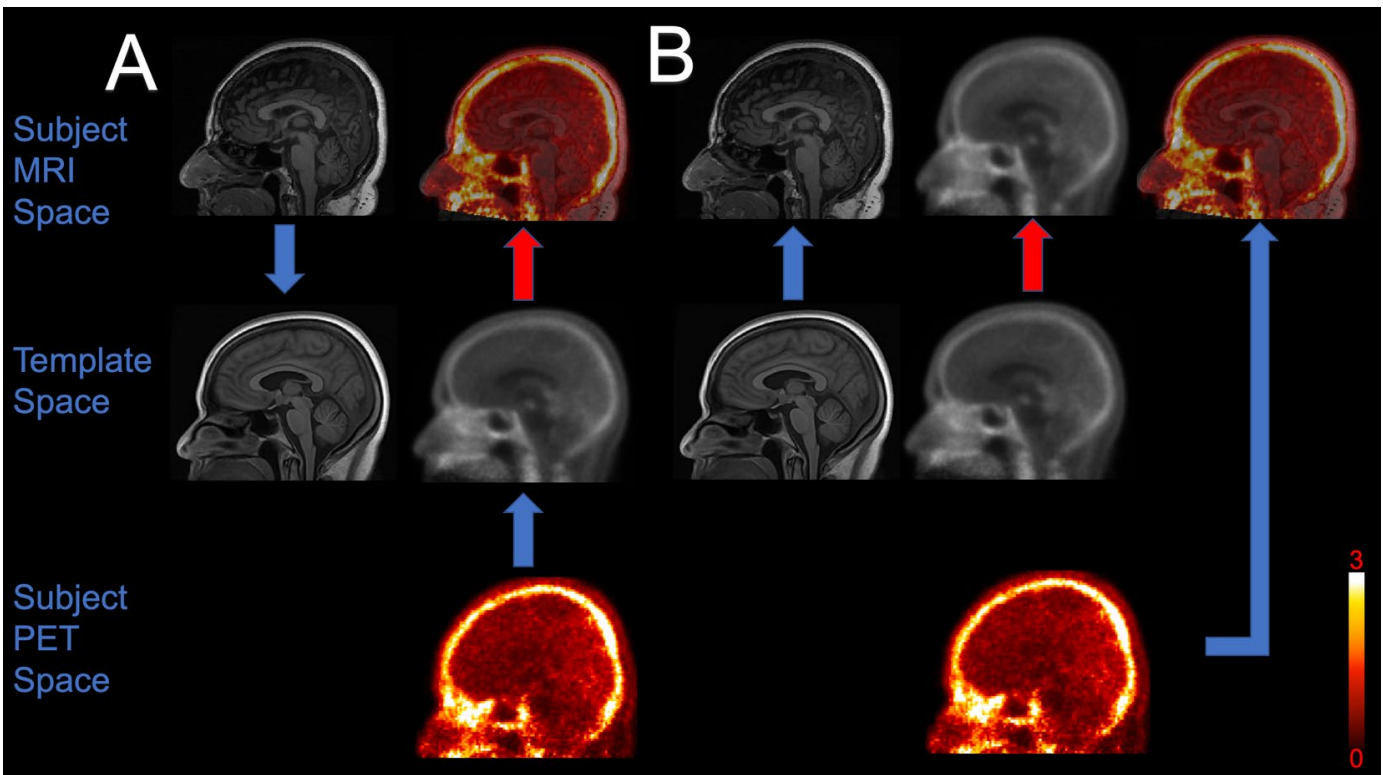

**Figure 4: Schema of Template aided Registration proposed here.** Blue arrows represent registration, while red arrows represent an applied transform. A. Template based registration (Template_1) passing the moving image (subject PET image) through template space. The PET and MRI image for each subject is first registered to their respective templates. We then apply inverse of the transform from MRI space to template space and apply this to the moving image. B. Alternative template based registration which avoids passing the moving image into template space (Template_2 and Template_3). First, we register the MRI template to the subject MRI, and then apply this transform to the PET template to create a template specific to each subject space. We then register the source image to this subject specific PET template.

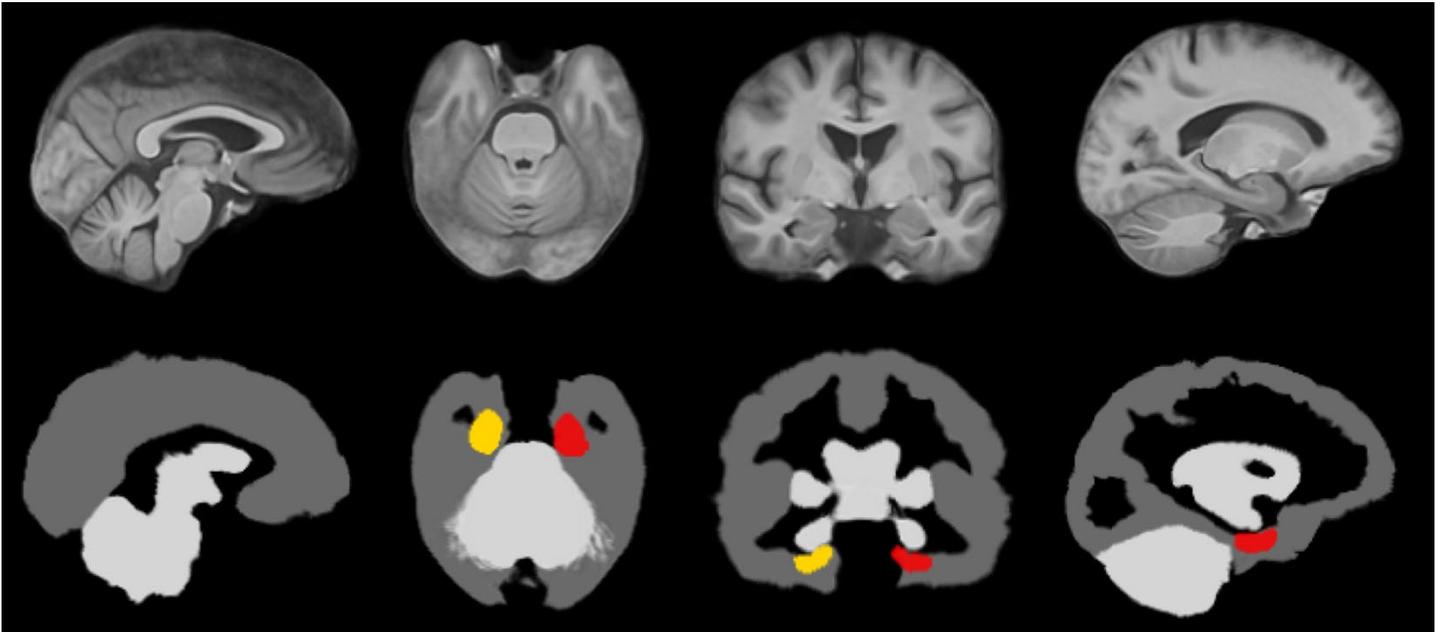

**Figure 5: MRI template (top row) with atlases (bottom row) we derived. Our first atlas segmented cortical (dark grey), subcortical (black), and deep brain and cerebellum (light grey). Our second atlas is a dilated and smoothed left (yellow) and right (red) entorhinal cortex.**

**Atlas Creation/Regional Uptake Calculation**

To test the utility of our template in the absence of subject MRI, we then used FreeSurfer (*29*) to create a segmentation of the MRI template based on the Desikan-Killiany Atlas (*30*). The template manually inspected/corrected onto which we created a smoothed and dilated bilateral entorhinal region of interest (Figure 5). We additionally also applied a segmentation based on an OASIS template (*31*) to segment cortical and subcortical volumes of interest (Figure 5), from which we calculated standardized uptake values from subject PET scans directly registered to template space. We then derived standardized uptake value ratios (SUVR) for our dilated entorhinal mask by dividing each uptake value by value derived from a cerebellar reference region. We then compared this to the uptake values we obtained using CT-MR based registration and spatial normalization to the PET template.

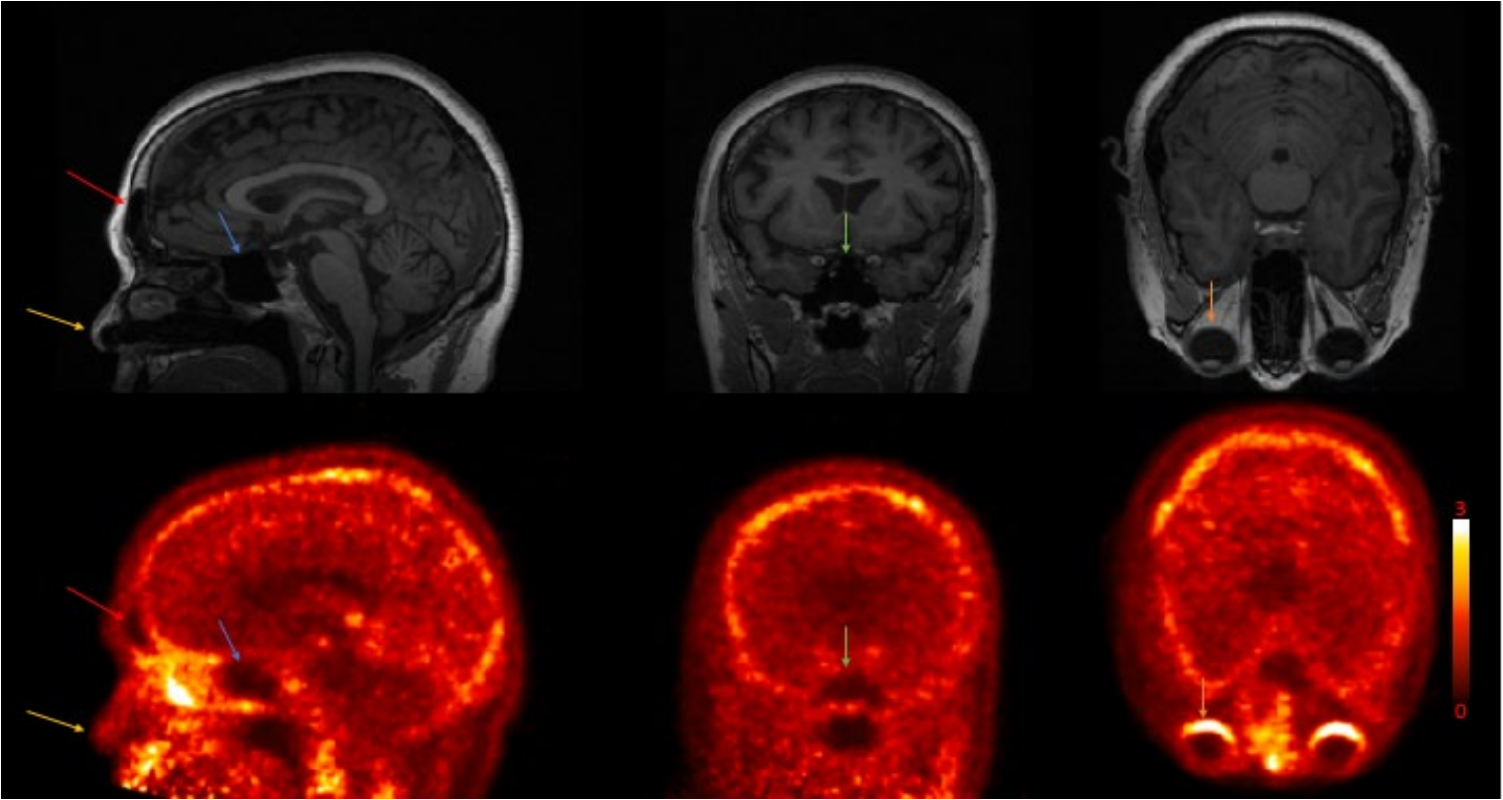

**Figure 6: Examples of landmarks used for evaluation of registration.** Seemingly paradoxically, we found that oftentimes the best fiducial landmarks for evaluation of registration seemed to be in areas of relatively decreased signal, such as the nose, sphenoid sinus, or frontal sinus.

**Performance Metrics/Statistical Analysis**

To evaluate these above described methods, we used Python (3.6) packages available in NiBabel (version 2.4.1, https://doi.org/10.5281/zenodo.3233118) and Numpy Version 1.17, (*32*) to load imaging files and analyze them respectively. We used packages available in Scipy to analyze transformation matrixes (*33,34*). For statistical analysis, we used R (Version 3.3.2). Example code for procedures (where appropriate) is at github.com/jszou/MK6240_coregistration.

The accuracy of registration using the CT scan approach was verified visually for each of the 60 subjects, paying close attention to correspondence areas of marked increased signal such as the eyes, cavernous sinus, and leptomeninges (see Figure 6 above). We then used this as the standard to which we compared the other co-registrations. We then masked each registered image with a brain mask derived from the brain extracted MRI to only evaluate brain areas.

To quantify the performance of the three above template based method as well as direct registration ($\dot{j}$), we performed a root mean square error analysis (defined as the root sum of the square of difference between two scans at each voxel for each 3D image of dimension 176x176x256), compared against the subject PET registered to MRI with the aid of subject CT as the gold standard ($\dot{i}$), given by:

$$\sum_{x=1}^{176}\sum_{y=1}^{176}\sum_{z=1}^{256}(i_{x,y,z} - j_{x,y,z})^2$$

For translational differences, we took *X*, *Y*, and *Z* translational components of the respective 3D transformation matrices A of transformation T with general form:

$$T(x) = Ax = \begin{bmatrix} 1 & 0 & 0 & X \\ 0 & 1 & 0 & Y \\ 0 & 0 & 1 & Z \\ 0 & 0 & 0 & 1 \end{bmatrix}\begin{bmatrix} x \\ y \\ z \\ 1 \end{bmatrix}$$

and found the respective sum of squares difference between CT aided coregistration (*i*) and the other registration procedures (*j*) $S_{translational}$ for each subject S, given by:

$$S_{translational} = \left\{ \sum_{C=k}^{k=X,Y,Z} (C_i - C_j)^2 \middle| j \in (direct, template\_1, template\_2, template\_3) \right\}$$

Similarly, for rotational differences, we took the sum of the square of differences of the euler angles for rotations on the X, Y, and Z axis. We found the euler angle $(\alpha, \beta, \gamma)$ of the euler vector along each axis X, Y, Z for the rotational matrix (without the translational component detailed above) of each tested registration procedures (A'$_j$) along with the rotational matrix of our standard (A'$_i$), defined as:

$$A' = X(\alpha)Y(\beta)Z(\gamma) = X_\alpha Y_\beta X_\gamma$$

$$= \begin{bmatrix} \cos(\beta) & -\cos(\gamma)\sin(\beta) & \sin(\beta)\cos(\gamma) \\ \cos(\alpha)\sin(\beta) & \cos(\alpha)\cos(\beta)\cos(\gamma) - \sin(alph)\sin(\gamma) & -\cos(\gamma)\sin(\alpha) - \cos(\alpha)\cos(\beta)beta)\sin(\gamma) \\ \sin(\alpha)\sin(\beta) & \cos(\alpha)\sin(\gamma) + \cos(\beta)\cos(\gamma)\sin(\alpha) & \cos(\alpha)\cos(\gamma) - \cos(\beta)\sin(\alpha)\sin(\gamma) \end{bmatrix}$$

With the set of differences $S_{euler}$ between CT aided coregistration (i) and other registration procedures (j) for each subject S, given by:

$$S_{euler} = \left\{ \sum_{C=k}^{k=\alpha,\beta,\gamma} (C_i - C_j)^2 \middle| j \in (direct, template\_1, template\_2, template\_3) \right\}$$

For statistical analysis, we performed a one-way ANOVA, and then pairwise t-tests to test the differences between errors for each of the four methods. We then re-performed this analysis with Bonferroni correction for multiple comparisons. For evaluation of our MR-less SUV derivation, we performed a one-way ANOVA to test the differences between SUVRs derived from each respective method, and performed correlation analysis to compare uptake values for each subject with each method. We plotted Bland-Altman plots to compare registration methods and SUV derivation methods.

For external validation of our registration techniques, we additionally tested each registration method on 30 holdout subjects whom we were able to acquire CT scans for in addition to their PET and MRI scans. We repeated the above validation testing for each subject.

## Results

|  | All controls (n= 201) | Template (n = 30) | p values |
|---|---|---|---|
| Age (mean, SD) | 65.8 +- 3.2 | 65.2 +- 3.5 | p=0.31 |
| Gender (M/F) | 65/136 | 13/17 | p=.81 |
| Race (White/Black/Hispanic) | 11/20/170 | 1/3/26 | p=.99 |
| Intracranial Volume (mm3) | 1375723± 145236 | 1384822 ± 149228 | p=.71 |

**Table 2:** Basic Demographic information for control population as well as 30 random, amyloid negative subjects.

Demographic data for all subjects and specifically for subjects included into the template are summarized in table 2. All basic demographic measures between the subject population and those randomly selected for the template were not significant, which included age (p = 0.31), gender (Chi-square p = 0.81), Race (chi-square p = 0.99) and intracranial volume (p = 0.71).

Overall, the four methods tested did not vary significantly from one another (p=0.511, F=0.884) in RMSE. Looking at direct comparisons, there were no significant differences in errors between methods (p's >0.3). Direct registration had a lower RMSE in the majority of cases in the original group evaluated (n=18/30, average = 0.0427, SD = 0.1095), while our alternative methods had the lowest RMSE in the rest.

Overall, translational errors did not vary significantly (p=0.228, F = 1.471). Errors between methods were not significantly different (p's > 0.1). Evaluating the translational portion of the transformation matrices of the aforementioned registrations, direct registration had the lowest error in 15 of 30 cases. Registration using the transformed template with mutual information had the smallest error in seven cases, and with correlation ratio seven cases.

Overall, there was no difference in rotational errors (p=0.577, F = 0.312). There were no significant pairwise differences between errors for any method (p's >0.1). When we looked at the angular difference (Euler vector) for each of the transformations, direct registration had the smallest angle in nine cases (average error = 0.21, SD = 0.29). The intermediate template had the smallest error in five cases (average error = 0.32, SD = 0.33). The transformed template with mutual information in six cases (average error = 0.221, SD = 0.291). The transformed template with correlation ratio in ten cases (average error = 0.204, SD = 0.257).

Results are summarized Table 3 below and Figure 6.

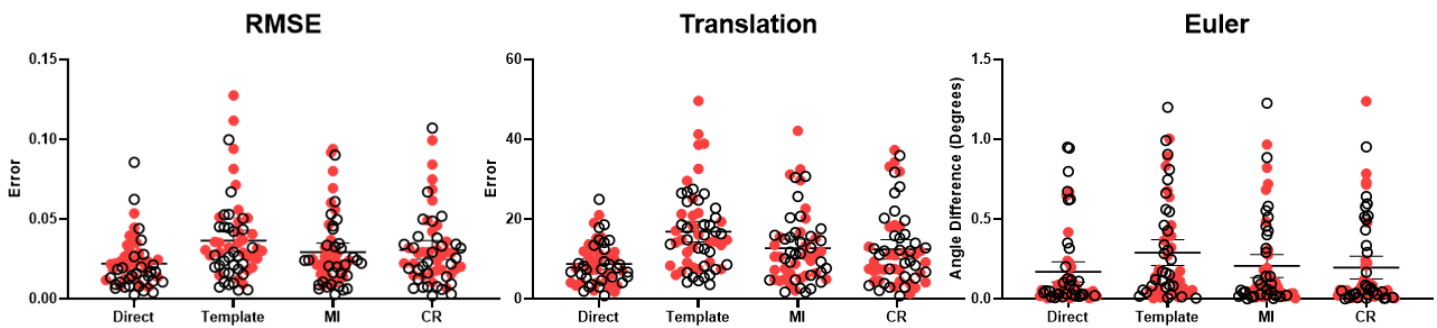

**Figure 6:** Scatter plots with mean and one standard deviation, showing errors for each of the metrics discussed. Red dots represent holdout subjects. We excluded the one outlier direct registration failure on this plot for purposes of formatting.

|  | RMSE |  | Translational Error |  | Rotational Error |  |
|---|---|---|---|---|---|---|
| **Method** | Avg (STD) | P value (effect size) | Avg (STD) | P value (effect size) | Avg (STD) | P value (effect size) |
| **Direct Registration** | 0.0427 (0.1095) |  | 42.1 (179.9) |  | 0.216 (0.294) | n/a |
| **Intermediate Template** | 0.0302 (0.0262) | 0.47 (0.157) | 14.9 (7.6) | 0.25 (0.216) | 0.323 (0.329) | 0.17 (0.344) |
| **Transformed Template (MI)** | 0.0257 (0.0207) | 0.33 (0.216) | 12.4 (7.7) | 0.21 (0.235) | 0.221 (0.291) | 0.95 (0.017) |
| Transformed Template (CR) | 0.0272 (0.0219) | 0.37 (0.196) | 12.9 (8.9) | 0.22 (0.231) | 0.204 (0.257) | 0.88 (0.044) |

**Table 3: Results for 30 template subjects.** P values represent the difference between respective errors of direct registration as compared to template based methods. Pairwise t-tests revealed no significant differences between methods.

**Holdout Set**

In the holdout set, direct registration had the lowest RMSE (17/30), translational error (17/30) and angular difference (14/30) in the majority of cases. Overall, the errors between the methods did not vary significantly for RMSE (p=0.296, F=1.121), translational error (p = 0.543, F = 0.376), or angular difference (p=0.956, F=0.003). There was a significant difference between direct registration and the intermediate template method in both RMSE (p<0.003, Cohen's D = .821) and translational error (p <$5\times10^{-5}$, Cohen's D = 1.13) which survived Bonferroni correction. There was a significant difference in translational error between the template method and transformed template method using correlation ratio (p < 0.01, Cohen's D = 0.64). There were no significant differences in error for any metric between methods. These are summarized in Table 4 below.

| | RMSE | | Translational Error | | Rotational Error | |
|---|---|---|---|---|---|---|
| Method | Avg (STD) | P value (effect size) | Avg (STD) | P value (effect size) | Avg (STD) | P value (effect size) |
| **Direct Registration** | 0.0402 (0.0609) | n/a | 65.42 (215.4) | n/a | 0.147 (0.211) | n/a |
| **Intermediate Template** | 0.0570 (0.0624) | 0.29 (0.272) | 75.14 (213.0) | 0.86 (0.045) | 0.292 (0.323) | 0.054 (0.531) |
| **Transformed Template (MI)** | 0.0470 (0.0614) | 0.67 (0.111) | 69.17 (214.1) | 0.95 (0.017) | 0.218 (0.286) | 0.344 (0.282) |
| **Transformed Template (CR)** | 0.0478 (0.0607) | 0.63 (0.125) | 68.38 (214.5) | 0.96 (0.0137) | 0.214 (0.317) | 0.368 (0.249) |

**Table 4: Results for 30 holdout subjects**. P values represent the difference between respective errors of direct registration as compared to template based methods. * = survives Bonferroni correction for multiple comparisons.

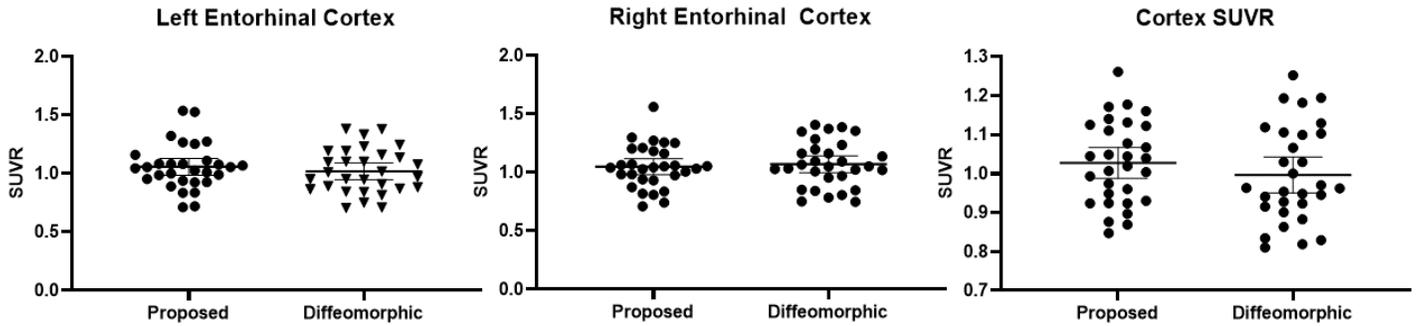

**Figure 7: Scatter plot showing SUVR for our proposed method and conventional diffeomorphic registration.** Bars show mean and one standard deviation.

**SUVRs**

SUVR values for each subject, using both our method and the diffeomorphic method, are reported above (Table 2). There were no significant differences between methods for any region of interest interrogated (p's >0.3, see Figure 7). Values derived for Cortical SUVR (R = 0.75 p= 1.6 x $10^{-6}$); and left (R = 0.83 p=1.4x$10^{-8}$) and right (R = 0.82, p = 2.8x$10^{-8}$) entorhinal cortex were all highly correlated (R = 0.95, p=2.2x$10^{-16}$) between these two methods. See Figure 8 below.

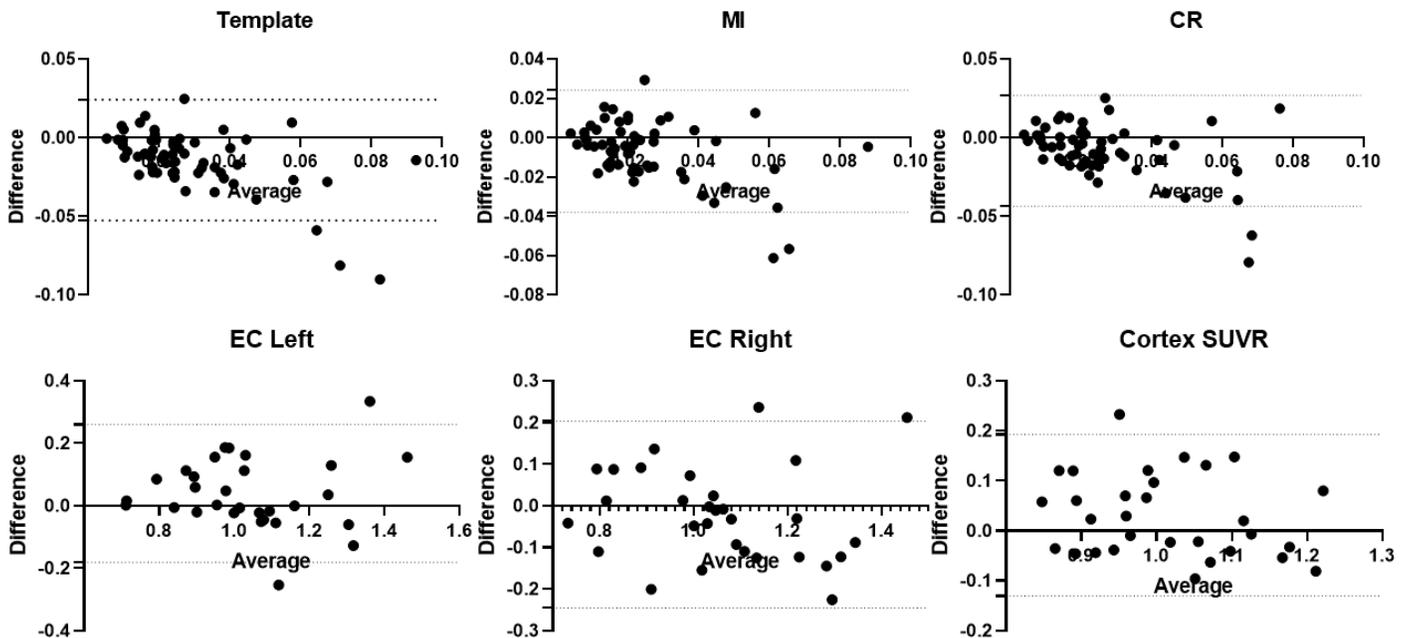

**Figure 8: Bland-Altmann Plots for registration RMSE (top row) and SUV quantification difference (bottom row) for methods vs. conventional registration.** Y-Axis represents difference between measures, while X-axis represents average of measures per subject. Dotted lines represent 95% confidence intervals.

## Discussion

We demonstrate here that conventional, direct registration of subject PET to subject MRI, though laden with potential pitfalls, is a valid approach for quantitative analysis of $^{18}$F-MK6240 assuming an adequate MRI is available. Additionally, we show that template based registration using a population derived template provides a feasible and valid alternative for the registration of $^{18}$F-MK6240 PET images to subject MRI images, especially in cases where direct registration may yield poor results and threaten to reduce the power of studies for which it can be burdensome or impossible to collect data. More importantly, in cases where a corresponding MRI scan is unavailable (or is inconsistent in quality or timing of acquisition) or there is a requirement for the processing of a scan for clinical evaluation or quantitative imaging, the ability to use only a subject's tau-PET may be crucial. This includes the application of this PET radioligand to non-research studies as well as facilitate the multi center sharing of imaging data, which has been reported previously with other tracers (*14,16*) but not to our knowledge with $^{18}$F-MK6240. This may also be important for the future clinical application of this radioligand, as the aging population are more likely to have implants, like aneurysm clips and pacemakers, that are of potentially unknown origin or contraindicated for MRI scanning, as is more common.

While prior studies have used multiple volumetric templates for a single image registration (*11,13*), we propose a single intermediary image volume, which we believe has the advantage of being a simpler model and thus requiring fewer pre-processing steps. This method may be beneficial with its ability to account for sources of signal intensity that may be related to yet-to-be-understood population specific signatures of off target binding or site-specific differences in PET image acquisition (*35*).

The poorer performance of template-based methods in the holdout group compared to testing in the original group from which the template was derived is not unexpected. This suggests that the creation of a study specific template may also be warranted for any site performing a large number of $^{18}$F-MK6240 scans, especially as aforementioned acquisition protocols and parameters may vary greatly per site. Additionally, while there are perhaps diminishing returns associated with creating larger template images, creation of a template from a larger cohort may help to reduce the variance in the average image generated. This may have an effect in positively biasing testing on the original cohort and negatively biasing testing on the holdout set. However, there is a tradeoff in a limited scan population of whom to include and holdout, as well as considerations of data leakage in evaluation with a smaller dataset. Additionally, incorporating a diffeomorphic transformation for this method may help account for individual anatomy more precisely, especially in the case of mild atrophy, but doing so would add significant computational cost and require further consideration with respect to areas of atrophy.

While, by most metrics we tested, direct registration of a subject PET image to MRI was a superior method to template based methodology, we believe that the premise, subject whole brain PET registered to whole brain MRI, is not ideal nor standard way to perform image registration, in part because images acquired at different time points may reflect markedly different anatomy (i.e. major changes in subject weight or recent surgery) or disease state. This is an especially salient point in the case of this relatively understudied tracer, where extracranial signal may serve as the dominant source of shared signal between modalities, and whose off-target kinetics may change depending on the subject or even the method of acquisition (full dynamic vs. window). This introduces an element of variance to the fidelity of registration which may be unpredictable, especially as there is no easy/practical way to know which areas drive registration for each subject. And, because these off-target sites often abut important regions, (aforementioned inferior cerebellum, leptomeninges, skull base and mesial temporal regions), whole brain registration may result in poor estimation.

A similar approach to SUVR quantification without the usage of MR has been previously attempted in the tau tracer $^{18}$F-AV-1451 (*14*), where multiple templates were tested including one derived from a principal components analysis, who noted exceptional performance of their template in mesial temporal areas, which are of obvious importance in the early detection of AD. Other past works have implemented a method using iterative generation of atlases which are subject specific (*36*), but these approaches, which generally require significant processing time for each subject, may be ill suited for clinical usage. Notably, we chose not to perform any formal spatial normalization with our MR-less technique, as this in theory is partially accounted for by the nature of the template we generated. Regardless, the application of recent advances to spatial normalization without MRI (*18,36,37*) is expected to improve the performance of our method, and may be the most accurate available method. Further efforts should be placed to explore the value of SUVR ROI accuracy relative to co-registration techniques (i.e. how our template ROI compares to a segmented ROI in patient MRI space) and considerations of partial volume correction (PVC)

As we have noted, we did not test either of our method on subjects noted to have clinical AD or imaging signs of AD (i.e. Braak staging patterns of radioligand deposition), as these scans generally do not have the same issue of signal intensity discordance which plague 'negative' scans. The use of this tracer in cognitively normal (or non-demented) individual, generally free of intracerebral uptake, presents a unique situation, where the earliest evidence of suspected pathology is differentially more difficult to evaluate and process. It is also not well known how sensitive this radioligand is for the detection of minute changes in disease progression at later stages, a time at which it may already be too late to intervene. Further work is needed to understand the role of tau-PET in early disease state, including MCI, as well as methods that focus on cerebral registration techniques.

There are some shortcomings to this study. Our population, a cohort of non-demented, largely female Hispanics, was not intended to reflect a larger population at risk for Alzheimer's disease or dementia. Being a post-hoc analysis, we were limited to using CT information rather than true fiducial landmarks (i.e. external markers on the patient) to use as our gold standard. We believe though that the attenuation CT that is frequently acquired in PET/CT machines at time of PET acquisition is underutilized for the purposes of analysis, and may provide the most accurate registration, both because (1) CT and MRI have more highly correlated structural density/intensity, and (2) the CT and PET are acquired at the same time and thus there are no concerns about differences in differences in patient disease state. The usage of CT for analysis may thus be warranted in the interrogation of small structures sensitive to registration errors (such as transentorhinal region), even if such structures are not visible on CT. This is of course not always possible, especially in large studies where acquiring and processing a CT scan for every subject may be unwarranted or unavailable (i.e. for PET only scanners). Additionally, we did not perform partial volume correction, as this usually requires the use of a subject MRI for purposes of segmentation (which would be unavailable in our proposed use case) and in theory would likely not improve the SUVR accuracy in target areas as we expect there to be little underlying signal within the subject image. Using a population derived template MRI (such as the one generated in this study) for purposes of initial segmentation in partial volume correction may be a viable alternative and warrants future investigation.

In conclusion, we evaluated registration methods, finding that currently used practices are comparable to existing methods with several important considerations. We also devised and evaluated an alternative method of registration as well as quantification using population derived intermediary templates in the absence of corresponding CT or MRI. Our method is a feasible alternative to conventional direct registration, especially when obvious structural or functional differences between images are present and provides an alternative framework for analyzing any PET radioligands for which there are no standard preprocessing approaches.

## Disclosures

Dr. Kreisl is a consultant for Cerveau Technologies. However, Cerveau was not involved in the design or execution of this study or in the interpretation of results.

Dr. Provenzano is a consultant for and has equity in Imij Technologies, an unrelated company, which was not involved in the design or execution of this study or the interpretation of results.

## Funding

Data collection was supported by the National Institutes of Health (NIH): grants R01AG050440, R01 AG055422, RF1AG051556, RF1AG051556-01S2, R01AG055299, K99AG065506 and K24AG045334. Partial support for data collection was provided by NIH grant UL1TR001873.

18F-MK6240 Registration 188.  Betthauser TJ, Cody KA, Zammit MD, et al. In Vivo Characterization and Quantification of Neurofibrillary Tau PET Radioligand 18F-MK-6240 in Humans from Alzheimer Disease Dementia to Young Controls. *J Nucl Med*. 2019;60:93-99.

9.  Pascoal TA, Shin M, Kang MS, et al. In vivo quantification of neurofibrillary tangles with [18 F] MK-6240. *Alzheimer's research & therapy*. 2018;10:74.

10. Gispert J, Pascau J, Reig S, et al. Influence of the normalization template on the outcome of statistical parametric mapping of PET scans. *Neuroimage*. 2003;19:601-612.

11. Ding L, Goshtasby A, Satter M. Volume image registration by template matching. *Image and Vision Computing*. 2001;19:821-832.

12. Lee H, Lee J, Kim N, Lyoo IK, Shin YG. Robust and fast shell registration in PET and MR/CT brain images. *Computers in Biology and Medicine*. 2009;39:961-977.

13. Schreibmann E, Xing L. Image registration with auto-mapped control volumes. *Medical Physics*. 2006;33:1165-1179.

14. Bourgeat P, Villemagne VL, Dore V, et al. PET-only 18F-AV1451 tau quantification. In: 2017 IEEE 14th International Symposium on Biomedical Imaging (ISBI 2017). ; 2017:1173-1176.

15. Bourgeat P, Villemagne VL, Dore V, et al. Comparison of MR-less PiB SUVR quantification methods. *Neurobiology of Aging*. 2015;36:S159-S166.

16. Edison P, Carter SF, Rinne JO, et al. Comparison of MRI based and PET template based approaches in the quantitative analysis of amyloid imaging with PIB-PET. *NeuroImage*. 2013;70:423-433.

17. Schöll M, Lockhart SN, Schonhaut DR, et al. PET imaging of tau deposition in the aging human brain. *Neuron*. 2016;89:971-982.

18. Alvén J, Heurling K, Smith R, et al. A Deep Learning Approach to MR-less Spatial Normalization for Tau PET Images. In: Shen D, Liu T, Peters TM, et al., eds. Medical Image Computing and Computer Assisted Intervention – MICCAI 2019. Lecture Notes in Computer Science. Cham: Springer International Publishing; 2019:355-363.

19. Ségonne F, Dale AM, Busa E, et al. A hybrid approach to the skull stripping problem in MRI. *NeuroImage*. 2004;22:1060-1075.

20. Fischl B, Salat DH, van der Kouwe AJW, et al. Sequence-independent segmentation of magnetic resonance images. *NeuroImage*. 2004;23:S69-S84.

21. Barthel H, Gertz H-J, Dresel S, et al. Cerebral amyloid-β PET with florbetaben (18F) in patients with Alzheimer's disease and healthy controls: a multicentre phase 2 diagnostic study. *Lancet Neurol*. 2011;10:424-435.

22. Seibyl J, Catafau AM, Barthel H, et al. Impact of Training Method on the Robustness of the Visual Assessment of 18F-Florbetaben PET Scans: Results from a Phase-3 Study. *J Nucl Med*. 2016;57:900-906.

23. Klein A, Tourville J. 101 Labeled Brain Images and a Consistent Human Cortical Labeling Protocol. *Front Neurosci*. 2012;6.